%% file: main.tex
\documentclass[twocolumn]{IEEEtran}

\usepackage{amsmath,amssymb,amsfonts,cuted,amsthm}
\usepackage[final]{graphicx}
\usepackage{psfrag}
\usepackage{epsfig}
\usepackage[numbers,sort&compress]{natbib}
\usepackage{array,booktabs}
\usepackage[nolist,nohyperlinks]{acronym}
\usepackage{ucs} 
\usepackage[utf8x]{inputenc}
\usepackage[dvipsnames]{xcolor}
\usepackage{tikz}
\usepackage{tkz-tab}
\usepackage{multirow}
\usepackage{latexsym}
\usepackage{mathrsfs}
\usepackage{siunitx}
\usepackage{tikz}
\usepackage{tikz-3dplot}
\usepackage{circuitikz}
\usepackage{subfigure}
\usepackage{float}
\usepackage{arydshln}
\usepackage{breqn}
\usepackage{lipsum}
\usepackage{relsize}
\usepackage{balance}

\usetikzlibrary{calc}

\makeatletter
\def\blfootnote{\xdef\@thefnmark{}\@footnotetext}
\makeatother

\definecolor{bluerevision}{RGB}{0,112,192}

\newcommand{\RS}{R\&S\textsuperscript{\textregistered}}

\begin{document}
\title{Enabling 6G Wireless Communications: UWB Characterization of Corridors within the H-Band}
\author{Juan E. Galeote-Cazorla, Alejandro Ramírez-Arroyo, \\ Mauricio Rodríguez, Reinaldo Valenzuela and Juan F. Valenzuela-Valdés}
\maketitle

\begin{abstract}
Future sixth-generation of wireless system is expected to provide data-rates in the order of 1\:Tbps and latencies below 1\:ms. Among others, one of the most promising strategies to meet these requirements is to operate at higher frequencies than millimeter wave bands: the THz bands. Nevertheless, because of the higher losses and the detriment of classical propagation mechanisms, deploying systems operating at these frequencies becomes a real challenge. Consequently, short-range scenarios are of special interest since these effects of THz bands can be managed. This work conducts an extensive campaign within corridors at frequencies within the H-band in the range from 250\:GHz to 330\:GHz. For the first time in literature, an ultra wideband of 80\:GHz is studied simultaneously. Large scale effects are assessed by estimating and modeling path gain. The path gain exponent varies between $-$2.1 and $-$1.6, which is explained by a guiding effect also observed at millimeter wave bands. Small scale effects are also evaluated in terms of parameters such as rice $K$-factor, root mean squared delay spread and coherence bandwidth. Additionally, an analytical approximation based on the classical $N$-rays model is proposed obtaining an accurate representation of the wireless channel which is coherent with the empirical analysis. The full analysis reveals the suitability of these THz bands for deploying point-to-point links due to the predominance of the line-of-sight contribution respect to the reflected components.
\end{abstract}
    

\blfootnote{\noindent This work has been supported by grant PID2024.157242OB.C44 funded by MCIN/AEI/10.13039/501100011033 and by ERDF/EU. It has also been supported by grants PID2020-112545RBC54 and PDC2023-145862-I00, funded by MCIN/AEI/10.13039/501100011033 and by ERDF/EU and the European Union NextGenerationEU/PRTR and grant DGP\_PIDI\_2024\_00736 by Junta de Andalucía; and in part by the predoctoral grant FPU22/03392 (\textit{Corresponding author: Juan E. Galeote-Cazorla}).}
    
\blfootnote{\noindent Juan E. Galeote-Cazorla and Juan F. Valenzuela-Valdés are with the Department of Signal Theory, Telematics and Communications, Research Centre for Information and Communication Technologies (CITIC-UGR), University of Granada, 18071, Granada, Spain (e-mails: juane@ugr.es; juanvalenzuela@ugr.es).}

\blfootnote{\noindent Alejandro Ramírez-Arroyo is with the Department of Electronic Systems, Aalborg University (AAU), 9220 Aalborg, Denmark (e-mail:araar@es.aau.dk).}

\blfootnote{\noindent Mauricio Rodriguez is with the School of Electrical Engineering, Pontificia Universidad Católica de Valparaíso, Valparaíso 2340025, Chile (e-mail: mauricio.rodriguez.g@pucv.cl).}

\blfootnote{\noindent Reinaldo Valenzuela is with Nokia Bell Labs, Murray Hill, NJ, USA \linebreak (e-mail: reinaldo.valenzuela@nokia-bell-labs.com).}
 
\section{Introduction}\label{sec:Introduction}
\IEEEPARstart{N}{}ext generation wireless networks are expected to provide data-rates in the order of 1\:Tbps, latencies below 1\:ms and serve hundreds of users/devices simultaneously \cite{ref:Saad2020,ref:Zhang2019}. In order to do so, several approaches are proposed in the literature: artificial intelligence (AI) as an active actor on the system core, reconfigurable intelligent surfaces (RISs) for passive beamforming, or orbital angular moment (OAM) communications to multiplex information in orthogonal modes~\cite{ref:IMT-2030}. Nevertheless, one of the most popular solutions is the design of systems operating at higher frequencies than millimeter wave (mmWave) or \mbox{sub-6\:GHz} bands \cite{ref:Akyildiz2022}. The most recent releases of the Third Generation Partnership Project (3GPP) proposes the use of the sub-THz and THz bands (frequencies from 100\:GHz to 10\:THz) for the future sixth generation (6G) \cite{ref:Serghiou2022,ref:Rappaport2019}. One of the greatest advantages respect to lower frequencies is the available spectrum. Whereas mmWave and sub-6\:GHz are saturated with a great amount of services operating simultaneously, the THz bands are practically empty providing bandwidths orders of magnitude higher \cite{ref:GRTHz002}. A great variety of scenarios are expected to operate at these bands, for example: body area networks (BAN), unmanned aerial vehicle (UAV) networks, industrial Internet of things (IIoT) or local area vehicular to everything (V2X) communications. In particular, most of short-range scenarios are expected to operate at the THz frequencies \cite{ref:GRTHz001,ref:GRTHz002,ref:Han2022}. 

However, propagation characteristics at the THz bands become a challenge since attenuation is substantially higher than in mmWave or sub-6\:GHz bands. Additionally, classical propagation mechanisms (transmission, reflection and diffraction) suffer of a considerable detriment \cite{ref:Rappaport2019}. Moreover, effects such as diffuse scattering become especially relevant due to the reduced wavelengths \cite{ref:Degli-Esposti2007}. In this context, channel modeling and characterization emerges as a key tool to properly assess these behaviors. This can be addressed from theoretical and/or empirical perspectives studying parameters of large and small scale as well as specific properties of the components propagated through the environment \cite{ref:Han2022}. The former includes the analysis of the received signal in terms of path gain, which can be assessed with empirical models or from a more statistical perspective with fading models \cite{ref:Rappaport2017,ref:Durgin2002,ref:Galeote2025}. In contrast, the latter approach focuses on topics such as the components identification or direction of arrival analysis \cite{ref:Fleury1999,ref:Ramirez2023,ref:Ramirez2024}.

The state-of-art is currently studying propagation channel within THz bands \cite{ref:Serghiou2022,ref:Han2022}. Because of the aforementioned particularities, extensive measurements campaigns and studies are needed to fully understand how the transmitted signals behave at these bands. For instance, in \cite{ref:Piesiewicz2007,ref:Jansen2011,ref:Wang2023} the authors measure and model scattering produced up to 1\:THz by different materials typically used in buildings. On the other hand, several works focuses on the suitability of THz links for urban microcells in terms of propagation characteristics at 142\:GHz and 159\:GHz \cite{ref:Xing2021,ref:Shakya2024,ref:Lee2023}. Scenarios such as drone-to-drone (D2D) or smart rail mobility environments are also assessed at 145\:GHz and 300\:GHz respectively \cite{ref:Abbasi2023,ref:Guan2021}. Conversely, works such as \cite{ref:He2017,ref:Tang2021,ref:Kim2015,ref:Fu2019} focus on ultra short-range scenarios for IIoT applications evaluating path loss at frequencies in the range from 220\:GHz to 340\:GHz. Lastly, measurements are also conducted at indoor scenarios such as office buildings and data centers at frequencies between 140\:GHz and 325\:GHz \cite{ref:Ju2021,ref:Priebe2013,ref:Cheng2020}.

Corridor scenarios are of special interest since they provide access to most of the rooms inside the buildings. The majority of studies published in literature focuses on narrow bands at sub-6\:GHz and mmWave frequencies. Particularly at sub-6\:GHz, in \cite{ref:Lafortune1990} an extensive measurement campaign is conducted at 900\:MHz, modeling path loss empirically. Several campaigns are presented in \cite{ref:Tarng1997,ref:Yang1998} at 900\:MHz, 1.8\:GHz and 2.44\:GHz, where the propagation channel is modeled with ray-tracing techniques. Path loss is assessed and modeled at 1\:GHz, 1.9\:GHz, 3.6\:GHz and 6\:GHz in \cite{ref:Alvarez2003,ref:Oestges2006,ref:Ciccognani2005}. More recently, RIS-assisted communications within corridors are evaluated at 2.6\:GHz in \cite{ref:Ren2024}. On the other hand, at frequencies from 7\:GHz to 15\:GHz, works such as \cite{ref:Moraitis2025,ref:Batalha2019,ref:Hui2016} focuses on path loss modeling in line-of-sight (LoS) and non-LoS (nLoS) within corridors and corners. Moving towards mmWave bands, path loss is also modeled at frequencies in the range from 22\:GHz to 60\:GHz in \cite{ref:Oyie2018,ref:Chizhik2020,ref:Diago-Mosquera2022,ref:Geng2009}. In addition to large scale analysis, studies such as \cite{ref:Shen2021,ref:Diba2021,ref:Cai2020,ref:Yue2019,ref:Geng2009_2,ref:Maccartney2015} also focuses on the analysis of small scale effects at frequencies ranging from 26\:GHz to 73\:GHz including the evaluation of parameters such as Ricean $K$-factor and root mean squared (RMS) delay spread. Unlike lower frequencies, works focused on the THz bands are scarcer in the literature. Path loss is analyzed and modeled in some campaigns at 159\:GHz and 300\:GHz \cite{ref:Lee2024,ref:Li2022,ref:Wang2022}. In \cite{ref:Takahashi2025}, the authors conducted a measurement campaign at 300\:GHz with a bandwidth of 8\:GHz where identify and model the multipath components (MPCs) arriving at the receiver (Rx) in different usage scenarios. Similarly, a directional analysis is performed in \cite{ref:Wang2025} to identify MPCs in corridor measurements between 215\:GHz and 225\:GHz. 

In this work, a measurement campaign is conducted within two different corridors at the H-band in the full range from 250\:GHz to 330\:GHz and modeled with an analytical approach. To the best of the authors' knowledge, for the first time an extensive characterization over 80\:GHz using an ultra wideband (UWB) channel sounder is conducted. The transmitter (Tx) is placed in a fixed position emulating an access point (AP) giving service along the corridors. From the obtained channel transfer functions (CTFs), the large and small scale effects are studied. Path gain is evaluated and modeled with the classical power law obtaining the gain exponent of corridors for each frequency of the band. On the other hand, parameters such as Ricean $K$-factor, RMS delay spread and coherence bandwidth are thoroughly analyzed in terms of the Tx-Rx distance. In addition, a physical model based on the classical $N$-rays theory is proposed to replicate this type of scenarios obtaining high accuracy with measurements. Overall, both the empirical results and the analytical model presented in this article will be useful for the deployment of short-range systems with large bandwidth requirements such as independent data-centers connected by corridors.

The rest of the article is organized as follows. Section II explains the measurement setup and the corridor scenarios. Section III presents an analytical model of the corridors based on the classical $N$-rays models. Section IV details the obtained results from the measurements in terms of large and small scale effects. In addition, a comparison of the empirical results with the $N$-rays model is performed. Finally, Section~V summarizes the conclusions and future lines.

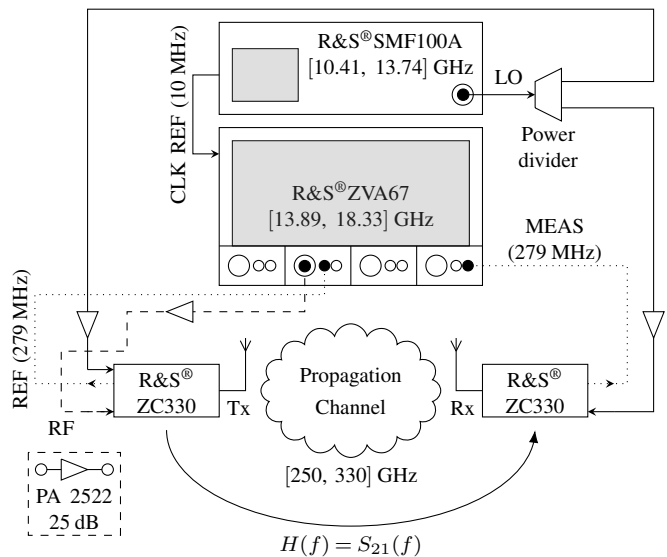
\begin{figure}[t!]
    \centering
    \begin{tikzpicture}[scale=0.7]
        \input{figures/channel_sounder}
    \end{tikzpicture}
    \caption{General scheme of the VNA-based frequency-domain channel sounder used for the measurements at the H-band.}
    \label{fig:channel_sounder}
\end{figure}

\section{Measurements Description}
\subsection{Channel Sounder}
The acquisition was performed using the frequency-domain channel sounder depicted in Fig.\:\ref{fig:channel_sounder}, which is based on a vector network analyzer (VNA). This sounder is designed to minimize the number of required cables and includes several amplifiers to increase the maximum measurable distance between Tx and Rx without compromising the dynamic range. To enable operation within the \mbox{H-band}, the VNA is connected to a pair of frequency converters at the Tx and Rx sides. The radiofrequency (RF) sounding signal is generated by the VNA at frequencies between 13.89\:GHz and 18.33\:GHz and fed to the Tx converter, which performs an up conversion by a factor of 18, resulting in a signal within the H-band that is transmitted and received by the antennas. Both converters are fed by a signal generator which provides a local oscillator~(LO) signal in the range from 10.41\:GHz to 13.74\:GHz. These LO signals are up converted by a factor of 24 and used to perform the down conversion of the received and transmitted signals to a fixed intermediate frequency (IF) of 279\:MHz. The down converted signals are then fed to the VNA through the measurement (MEAS) and reference (REF) ports, respectively, to compute the transmission parameter $S_{21}$. At each frequency $f$, this parameter is equivalent to the CTF; that is, $H(f)$\:$=$\:$S_{21}(f)$. Finally, the synchronization between Tx and Rx is ensured by the clock reference (CLK REF) of 10\:MHz provided by the signal generator to the VNA, as well as the power divider used to distribute the LO signals to both frequency converters. 

The VNA model used for the measurements was the ZVA67, which operates at frequencies up to 67\:GHz. The signal generator was the SMF100A model, capable of generating output signals up to 22\:GHz. The frequency converters were the ZC330 models, designed to operate within the full H-band and equipped with WR03 standard waveguide outputs. The typical output power is $-$7\:dBm. All devices were manufactured by Rohde \& Schwarz\textsuperscript{\textregistered}. On the other hand, the low-noise amplifiers were PA 2522 model from Langer EMV, which provide 25\:dB within the operation band. The power divider model was the PWD-2W-2G-18G-10W-Sf from AA MCS. Lastly, the Tx and Rx antennas were standard gain horns (SGHs), model \#32240 from Flann, designed for H-band operation. At the midband frequency, they provide a gain of 20\:dB and half power beamwidths (HPBWs) of 16.5º at both the H-plane and E-plane. Therefore, the effective isotropic radiated power (EIRP) is about 13\:dBm.

Finally, the channel sounder was configured to perform UWB measurements in the range from 250\:GHz to 330\:GHz with a sampling step $\Delta f$\:$=$\:10\:MHz ---that is, a bandwidth $\mathrm{BW}$\:$=$\:80\:GHz sampled with 8,001 frequency points per CTF. The intermediate frequency bandwidth~(IFBW) was set to 100\:Hz to balance the trade-off between time sweep and dynamic range. Calibration was performed using the normalized forward method, which sets the calibration planes at the output waveguides of the converters. As a result, the antenna gain patterns effects are part of the measured~CTF. The mentioned configuration yielded a noise floor of $-$111.5\:dB. All parameters are summarized in Table \ref{tab:channel_sounder_parameters}, including the delay resolution $\Delta\tau$\:$=$\:$1/\mathrm{BW}$\:$=$\:12.5\:ps and the maximum excess delay $\tau_{\max}$\:$=$\:($N$\:$-$\:1)\:$\Delta\tau$\:$=$\:100\:ns.

\begin{table}[t!]
    \centering
    \caption{VNA-based frequency-domain \\ channel sounder parameters}
    \label{tab:channel_sounder_parameters}
    \renewcommand{\arraystretch}{1.25}
    \begin{tabular}{|c|c|c|}
        \hline
        \multirow{7}{*}{\textbf{VNA}} & \textbf{Frequency range} & 250\:GHz to 330\:GHz \\ \cline{2-3}
        & \textbf{Meas. bandwidth [$\rm BW$]} & 80\:GHz \\ \cline{2-3}
        & \textbf{Frequency step [$\Delta f$]} & 10\:MHz \\ \cline{2-3}
        & \textbf{Number of points [$N$]} & 8,001 \\ \cline{2-3}
        & \textbf{Delay resolution [$\Delta \tau$]} & 12.5\:ps \\ \cline{2-3}
        & \textbf{Max. excess delay [$\tau_{\max}$]} & 100\:ns \\ \cline{2-3}
        & \textbf{Noise floor} & $-$111.5\:dB \\ \cline{2-3}
        & \textbf{IFBW} & 100\:Hz \\ \hline\hline
        \multirow{3}{*}{\textbf{Antenna}} & \textbf{EIRP} & 13\:dBm \\ \cline{2-3}
        & \textbf{HPBW} & 16.5º \\ \cline{2-3}
        & \textbf{Polarization} & Vertical \\ \hline\hline
        \multicolumn{2}{|c|}{\textbf{Dynamic range}} & 47.4\:dB \\ \hline
    \end{tabular}
\end{table}

\begin{figure}[t!]
    \centering
    \includegraphics{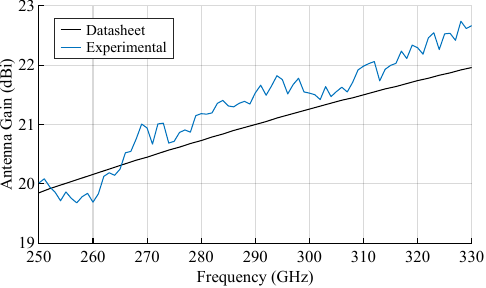}
    \caption{Maximum antenna gain as a function of frequency.}
    \label{fig:antenna_gain}
\end{figure}

\begin{figure}[t!]
    \centering
    \subfigure[H-plane.]{\includegraphics{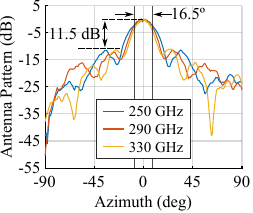}}
    \subfigure[E-plane.]{\includegraphics{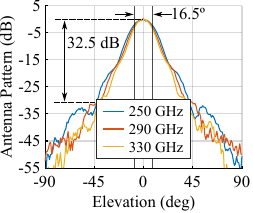}}
    \caption{Antenna gain patterns at the start, mid and end frequencies.}
    \label{fig:antenna_pattern}
\end{figure}

Since the antenna gain patterns are part of the measured propagation channel, an accurate characterization is needed to properly understand and model the performed measurements. In order to do so, the SGH antennas have been measured in the anechoic chamber located at the facilities of the Smart and Wireless Applications and Technologies (SWAT) group. Particularly, applying the well-known two-antenna method~\cite{ref:IEEEStd149}, the antenna gain values are estimated as a function of frequency and compared with the datasheet (see Fig.\:\ref{fig:antenna_gain}) obtaining mean differences of 0.4\:dB. On the other hand, the antenna gain patterns have been also characterized along the full band from 250\:GHz to 330\:GHz in both H-plane and E-plane (see Fig.\:\ref{fig:antenna_pattern}). As expected, the HPBW values are in the interval (16.5\:$\pm$\:1.7)º within all the frequency range. Nevertheless, it is noteworthy that the H-plane pattern present higher secondary lobes than the E-plane pattern, which can influence the measurement campaign results. In particular, these secondary lobes are 11.5\:dB and 32.5\:dB lower than the main lobe of the antenna respectively in H-plane and E-plane.

\subsection{Corridor Scenarios}
Measurements were conducted at two corridor scenarios located at the facilities of the Research Centre for Information and Communication Technologies (CITIC-UGR) and the Companies Centre for Information and Communication Technologies (CETIC-UGR), both affiliated with University of Granada (UGR). The corridor at CITIC-UGR has a width of 2.00\:m width, whereas the one at CETIC-UGR is 1.80\:m wide; both have a height of 2.65\:m and the over-ceiling gap is of~0.4\:m. The walls are constructed from concrete and are covered with plasterboard, plastic paint, plywood cabinets and doors. The floor consists of polished stone slabs, and the ceiling is composed of plasterboard tiles.

\begin{figure*}[t]
    \centering
    \includegraphics{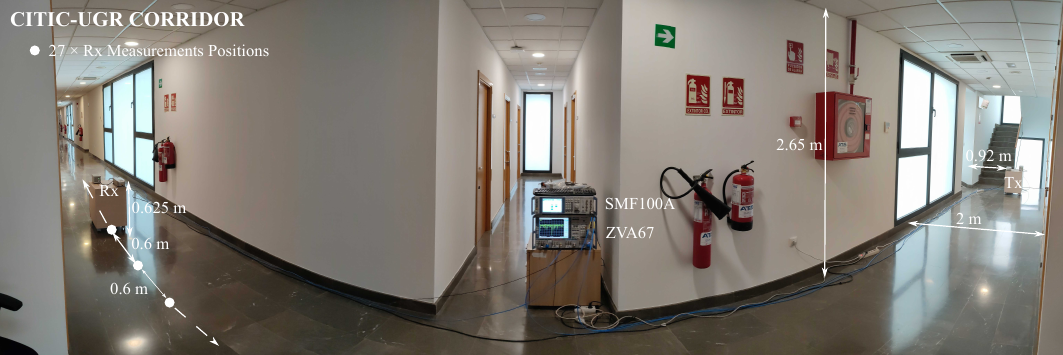}
    \caption{Panoramic view of the experimental setup in the CITIC-UGR corridor. Each white dot represents a Rx position spaced at intervals of 0.6\:m.}
    \label{fig:CITIC_setup}
\end{figure*}

\begin{figure}[t]
    \centering
    \includegraphics{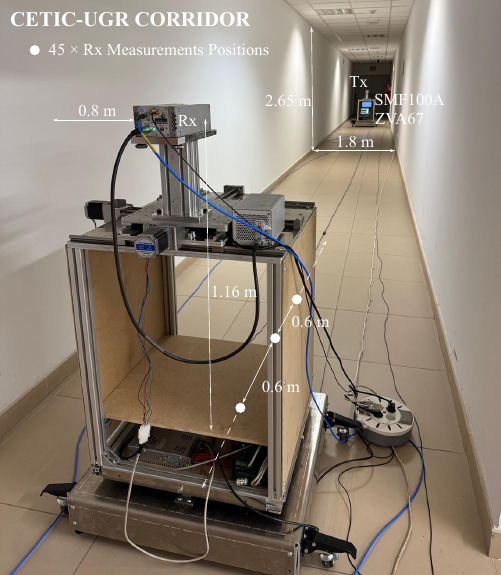}
    \caption{Picture of the experimental setup in the CETIC-UGR corridor. The Rx positions are represented by dots spaced at intervals of 0.6\:m.}
    \label{fig:CETIC_setup}
\end{figure}

Using the aforementioned sounder configuration, the experimental setup was defined as follows. The Tx was positioned at the end of each corridor, emulating an AP. In the CITIC-UGR scenario (see Fig.\:\ref{fig:CITIC_setup}), the Rx was aligned with the Tx at distances ranging from 0.6\:m to 16.2\:m, spaced at 0.6\:m intervals ---that is, a total number of 27 measurement locations. Both Tx and Rx were at a height of 62.5\:cm respect to the floor. On the other hand, in the CETIC-UGR scenario (see Fig.\:\ref{fig:CETIC_setup}), the Rx was also aligned with the Tx but at distances ranging from 1.2\:m to 27.6\:m. The adjacent locations were spaced 0.6\:m, yielding to 45 measurement positions. The height of both Tx and Rx was 116\:cm. Additionally, both Tx and Rx are separated 8\:cm and 10\:cm from the central axis of the corridors in CITIC-UGR and CETIC-UGR respectively. This will be relevant for the generalized model presented in the next section. Finally, the dynamic range for the largest Tx-Rx distance in CETIC-UGR was 47.4\:dB (see Table \ref{tab:channel_sounder_parameters}), which is adequate to properly capture both small-scale and large-scale effects.
 
\section{N-rays Corridor Model}
A simple way to model the propagation environment of this work is through the application of an $N$-rays model \cite{ref:Goldsmith2005}. This model describes the CTF as the sum of $N$ distinct contributions (or rays) arriving at the receiver via different paths. In this case, these contributions include the LoS component and the reflections from the walls, floor and ceiling, with an arbitrary number of bounces $n$. Therefore, the CTF for a Tx-Rx distance $d$ and frequency $f$ is computed as
\begin{equation}\label{eq:N-ray_model}
    H(f,d) = H_{\rm LoS}(f,d) + \sum_{i}H_i(f,d),
\end{equation}
where the term $H_{\rm LoS}(f,d)$ represents the LoS contribution and $H_i(f,d)$ accounts for the $i$-th reflected component arriving at the receiver. For the sake of simplicity, scattering contributions due to the presence of elements comparable to the wavelength (e.g. roughness of the walls) are not considered since they are weaker than the LoS and reflected components \cite{ref:Degli-Esposti2007}.

The LoS component of the CTF is computed directly using the Friis formula as follows \cite{ref:Friis1946}:
\begin{equation}\label{eq:LoS_CTF}
    H_{\rm LoS}(f,d) = G_{Tx}^{\max}(f)G_{Rx}^{\max}(f)\frac{\lambda}{4\pi d}\exp\left(-j\frac{2\pi d}{\lambda}\right),
\end{equation}
where $\lambda$ denotes the wavelength, and $G_{Tx}^{\max}(f)$ and $G_{Rx}^{\max}(f)$ represent the maximum antenna gains of the Tx and Rx respectively. Similarly, the $i$-th reflected component of the CTF can be computed as
\begin{equation}\label{eq:reflection_CTF}
    H_i(f,d) = \gamma_i(f)G_{Tx}^{(i)}(f)G_{Rx}^{(i)}(f)\frac{\lambda}{4\pi r_i}\exp\left(-j\frac{2\pi r_i}{\lambda}\right).
\end{equation}
In this case, $G_{Tx}^{(i)}(f)$ and $G_{Rx}^{(i)}$ represent the antenna gains of the Tx and Rx at the directions of departure and arrival respectively, and $r_i$ denotes the distance traveled by the ray along the followed path. Additionally, a factor $\gamma_i(f)$ is included to model the effects of the reflections: additional attenuation due to the transmitted waves through the walls and the corresponding change of phase.

Geometrically, an ideal corridor can be conceived as a rectangular cavity with dimensions $w$\:$\times$\:$h$\footnote{In the following, the dimension $w$ refers to the width of the corridor and $h$ represents its height from the floor to the ceiling.} that extends to the infinity without any imperfection in its surfaces (e.g. holes, doors, cabinets or windows). Resembling the experimental setup described in the previous section, the Tx can be placed at an arbitrary position of this corridor and the Rx shall be aligned with it at a distance $d$. Additionally, they are separated from the central axis of the corridor $\Delta w$ and $\Delta h$ in the horizontal and vertical directions respectively.

\begin{figure}[t!]
    \centering
    \subfigure[Odd number of bounces, $n$\:$=$\:3.]{
    \begin{tikzpicture}[scale=0.975]
        \input{figures/nray_model_odd.tex}
    \end{tikzpicture}
    }
    \par
    \subfigure[Even number of bounces, $n$\:$=$\:2.]{
    \begin{tikzpicture}[scale=0.975]
        \input{figures/nray_model_even.tex}
    \end{tikzpicture}
    }
    \caption{Vertical section of an ideal corridor with Tx and Rx aligned at a distance $d$. The LoS component is depicted with a gray line, and the two possible reflections are represented with the blue and orange lines.}
    \label{fig:N-ray_model}
\end{figure}
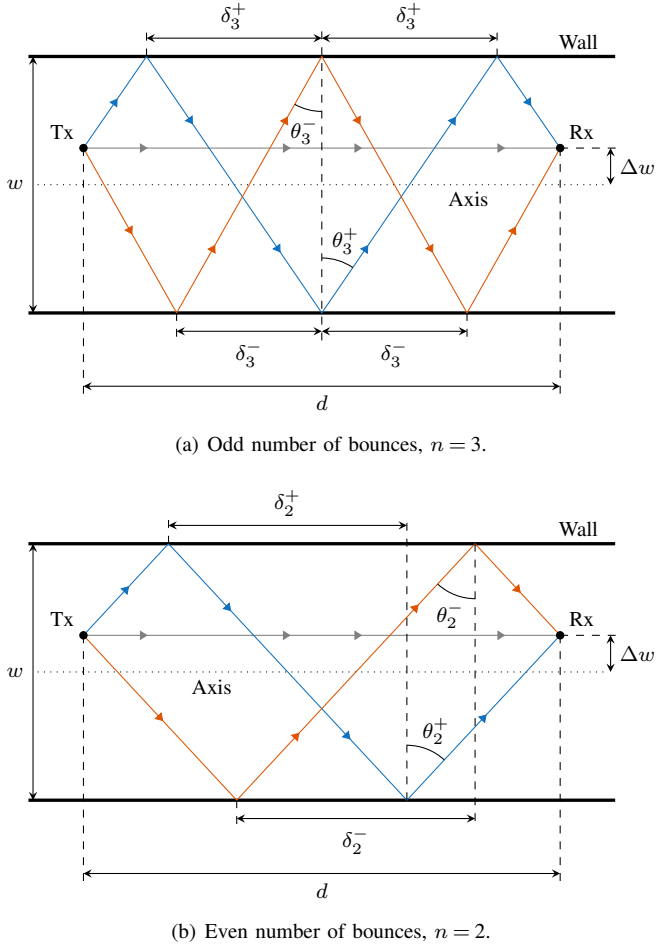

A sketch of the horizontal section of the ideal corridor is shown in Fig.\:\ref{fig:N-ray_model}, including both the Tx and Rx positions as well as the LoS component (gray line). For each number of bounces $n$, two different reflections appear (blue and orange lines), one for each wall of the corridor. It can be proven that the distance $\delta_n$ between two consecutive bounces is related to the Tx-Rx distance $d$ by the following expression:
\begin{equation}\label{eq:bounces_distance}
    \frac{d}{\delta_n^\pm} = n \pm [(-1)^n - 1]\frac{\Delta w}{w}.
\end{equation}
The solutions with the plus and minus signs correspond to reflections that begin bouncing from the two opposite walls. It is worth noting that for even values of $n$, both solutions become symmetrical. Finally, the distance $d_n$ traveled by each ray along its path is related to $\delta_n$ by the equation
\begin{equation}\label{eq:reflections_distance}
    \frac{d_n^\pm}{d} = \sqrt{1 + \left(\frac{w}{\delta_n^\pm}\right)^2}.
\end{equation}
Similarly, the angle of reflection $\theta_n^\pm$ is computed as
\begin{equation}\label{eq:reflections_angle}
    \theta_n^\pm = \tan^{-1}\left(\frac{\delta_n^\pm}{w}\right).
\end{equation}
All of these results are also valid for the vertical section of the corridor by replacing $w$ and $\Delta w$ with $h$ and $\Delta h$ respectively. Therefore, four possible components are computed using this method for each number of bounces $n$.

\begin{figure*}[t!]
    \centering
    \subfigure[CITIC-UGR (empirical).]{\includegraphics{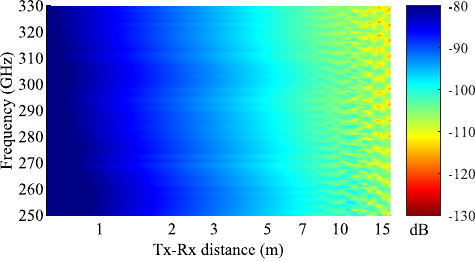}} \hspace{0.05\linewidth}
    \setcounter{subfigure}{2}
    \subfigure[CETIC-UGR (empirical).]{\includegraphics{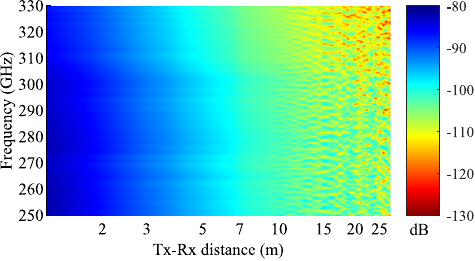}} 
    \setcounter{subfigure}{1}
    \subfigure[CITIC-UGR ($N$-rays model)]{\includegraphics{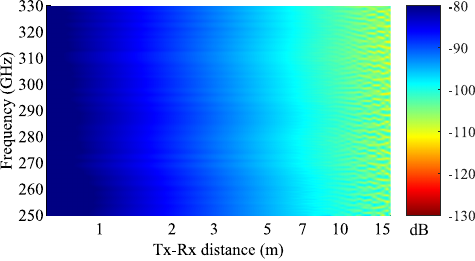}} \hspace{0.05\linewidth}
    \setcounter{subfigure}{3}
    \subfigure[CETIC-UGR ($N$-rays model).]{\includegraphics{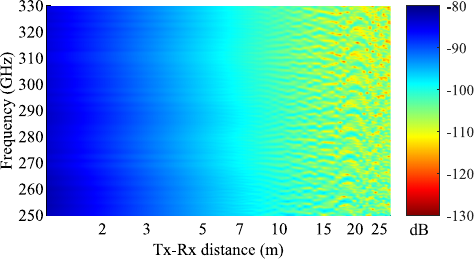}}
    \caption{Estimated path gain (in decibels) from the CTFs for the measurements and the $N$-rays models in CITIC-UGR and CETIC-UGR corridors.}
    \label{fig:path-gain}
\end{figure*}

The effect of reflections can be evaluated by applying the well-known Fresnel equations\footnote{In the particular scenario of this work, the incident wave propagates through air, whose relative permittivity can be approximated as $\varepsilon_r$\:$=$\:1.} \cite{ref:Molisch2011}. In order to do so, the angle of refraction $\alpha_n(f)$ must first be computed: 
\begin{equation}\label{eq:refraction_angle}
    \alpha_n(f) = \sin^{-1}\left(\frac{\sin\theta_n}{\sqrt{\varepsilon_r(f)}}\right),
\end{equation}
where $\varepsilon_r$ denotes the relative permittivity of the refracted medium. The permittivity of the different materials is obtained from the Recommendation ITU-R P.2040-4 \cite{ref:ITU-R-P.2040}. Specifically, the walls are modeled as ``plasterboard'', the floor as ``concrete'' and the ceiling as ``ceiling board''; whose $\varepsilon_r$ values are 2.56, 5.17 and 1.52 respectively. Then, the reflection coefficients $R^{(n)}(f)$ for transverse electric (TE) and transverse magnetic~(TM) incident waves are given by
\begin{equation}\label{eq:TE_reflection}
    R_{\rm TE}^{(n)}(f) = \frac{\cos\theta_n - \sqrt{\varepsilon_r(f)}\cos\alpha_n(f)}{\cos\theta_n + \sqrt{\varepsilon_r(f)}\cos\alpha_n(f)},
\end{equation}
\begin{equation}\label{eq:TM_reflection}
    R_{\rm TM}^{(n)}(f) = \frac{\cos\alpha_n(f) - \sqrt{\varepsilon_r(f)}\cos\theta_n}{\cos\alpha_n(f)+\sqrt{\varepsilon_r(f)}\cos\theta_n}.
\end{equation}
Since the horn antennas used in the experimental setup generate a vertically polarized waves, TE incidence is assumed for reflections from the walls and TM incidence for reflections from the floor and ceiling. Lastly, the factor $\gamma^{(n)}(f)$ is computed as the $n$-th power of the corresponding reflection coefficient, accounting for all bounces before reaching the Rx:
\begin{equation}\label{eq:reflection_factor}
    \gamma^{(n)}(f) = \left[R^{(n)}(f)\right]^n.
\end{equation}

The measurements explained in the previous section have been replicated with this model for a maximum number of bounces $N_b$\:$=$\:6. Thus, the model accounts for 4\:$\times$\:$N_b$\:$=$\:24 different rays arriving at the Rx, which ensures that the obtained CTFs have converged with changes respect to lower number of bounces in the order of 10\textsuperscript{$-$6} (linear scale). Intuitively, for larger Tx-Rx distances, a higher $N_b$ is needed to reach convergence in the CTFs since the angles of reflection are wider [see eqs. (4)--(6)]. This fact facilitates the alignment between the reflected rays and the antenna beams.

\section{Corridors Characterization \\ and Comparison with N-rays Model}

\subsection{Path Gain Analysis and Modeling}
Path gain is one of the key metrics of the wireless channel since it accounts for the attenuation of the signal propagating within the corridor. For LoS links, this parameter can be estimated from the CTF by subtracting the fast-fading effects and the antenna gain \cite{ref:Chizhik2020}. In narrowband measurements, the former can be addressed applying a moving average in distance with a window size between 20$\lambda$ and 40$\lambda$, with $\lambda$ equal to the wavelength \cite{ref:Lee1985}. However, that is not possible for this case since the spacing between adjacent measurements is $\Delta d$\:$=$\:0.6\:m, which is larger than the wavelengths in the frequency range. Nevertheless, it can be demonstrated (see Appendix A) that an equivalent criteria for a Tx-Rx distance $d$ is to apply a moving average in frequency with a window size given by
\begin{equation}\label{eq:window_size}
    L(d) = M\left\lceil\frac{d_{\rm max}}{d}\right\rceil + 1,
\end{equation}
where $d_{\rm max}$\:$=$\:$c/\Delta f$\:$=$\:30\:m is the maximum observation distance for the delay domain and $M$\:$\in$\:$\mathbb{N}$ such as 20\:$\leq$\:$M$\:$\leq$\:40. In this case, a value $M$\:$=$\:40 has been chosen\footnote{For the lowest distance ($d$\:$=$\:0.6\:m), a window size $L$\:$=$\:2,000 is obtained. Conversely, for the largest distance ($d$\:$=$\:27.6\:m), the size is $L$\:$=$\:45.}. Then, the windowed frequency response $|H_w(f,d)|$ is obtained as 
\begin{equation}\label{eq:remove_fast_fading}
    |H_w(f,d)| = \frac{1}{L(d)}\sum_{n=-\lfloor L(d)/2\rfloor}^{\lfloor L(d)/2\rfloor}|H(f-n\Delta f,d)|.
\end{equation}
Finally, the effects of the maximum antenna gain (see Fig.\:\ref{fig:antenna_gain}) can be removed to estimate the path gain:
\begin{equation}\label{eq:path-loss}
    \mathrm{PG}(f,d) = \frac{|H_w(f,d)|}{G_{Tx}^{\max}(f)G_{Rx}^{\max}(f)}.
\end{equation}

\begin{figure*}[t!]
    \centering
    \subfigure[Path gain exponent.]{\includegraphics{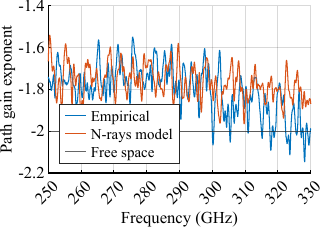}} \hfill
    \subfigure[1-meter intercept.]{\includegraphics{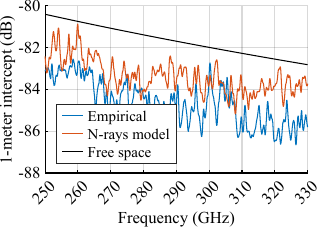}} \hfill
    \subfigure[Standard deviation.]{\includegraphics{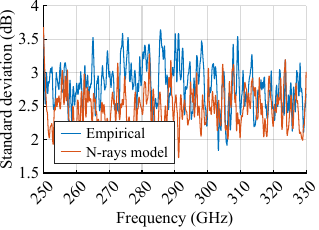}}
    \caption{Parameters obtained for the least squares minimization with the power law model as a function of frequency. The results of fitting the empirical samples and the $N$-rays model simulation are depicted in blue and orange. The free space propagation values are given as reference plotted with a black line.}
    \label{fig:fitting}
\end{figure*}

The obtained results for path gain from the experimental and the $N$-rays models are presented in Fig.\:\ref{fig:path-gain}. In both cases, similar behaviors are observed, which indicates a good concordance between the geometrical model and the empirical samples. In general, magnitude varies in the range defined between $-$130\:dB and $-$80\:dB. Particularly, at each Tx-Rx distance, path gain reduces with frequency due to the smaller wavelength. Differences in the order of 3\:dB are observed between 250\:GHz and 330\:GHz, which are close to the value of 2.4\:dB expected in free space. On the other hand, for each frequency, a clear linear trend is observed for distances $d$\:$\lesssim$\:7.2\:m. In contrast, for $d$\:$\gtrsim$\:7.2\:m, interference patterns appears in both frequency and distance axes with decays down to $-$130\:dB. These patterns are more noticeable as distance increases due to the increase on the received components. In addition, as frequency increases, the constructive/destructive interferences are more pronounced leading to deeper fadings.

In the literature, path gain is usually described by classical models such as close-in (CI) or alpha-beta-gamma (ABG)~\cite{ref:Rappaport2017}. Nevertheless, one of the most popular approaches for narrowband measurements is to fit the path gain with a power law model. For wideband measurements, this model can be generalized assuming a dependence on frequency in the constitutive parameters. Then, the expression of the general power law is given by
\begin{equation}
    \left.\widehat{\mathrm{PG}}(f,d)\right|_{\rm dB} = A(f) + 10n(f)\log_{10}\left(\frac{d}{d_0}\right) + X_{\sigma(f)},
\end{equation}
where $d_0$\:$=$\:1\:m is the reference distance, $n(f)$ is the path gain exponent, $A(f)$ represents the 1-meter intercept, $\widehat{\rm PG}(f,d)$ is the estimated path gain and $X_{\sigma(f)}$ is a zero-mean gaussian random variable (RV) with variance $\sigma^2(f)$. In order to fit empirical samples of path gain with the model, a least squares minimization is performed. Therefore
\begin{equation}\label{eq:fitting}
    \sigma^2(f) = \min_{\{n(f),A(f)\}}\sum_k\left[\mathrm{PG}(f,d_k) - \mathrm{\widehat{PG}}(f,d_k)\right]^2,
\end{equation}
with $k$ iterating over all the distance samples and path gain expressed in decibels.

A fitting with both CITIC-UGR and CETIC-UGR samples has been performed following \eqref{eq:fitting}. The obtained results from the empirical and simulated data with the $N$-rays model are depicted in Fig.\:\ref{fig:fitting}, and the expected values for free space propagation are given as reference. Fig.\:\ref{fig:fitting}(a) represents the values of $n(f)$, which takes mean values of $-$1.80 and $-$1.77 for the empirical samples and the $N$-rays model respectively. For frequencies $f$\:$\lesssim$\:300\:GHz, in both cases the path gain exponent varies between $-$1.9 and $-$1.6. In \cite{ref:Wang2025}, similar values are observed at 300\:GHz for narrow measurements. This behavior has also been observed at mmWave bands, where these values are related to a guidance effect within the corridor \cite{ref:Oyie2018,ref:Chizhik2020,ref:Diago-Mosquera2022,ref:Maccartney2015}. However, for $f$\:$\gtrsim$\:300\:GHz, the trend on the empirical results changes slightly approaching to the free space value of $-$2. This phenomena might be a consequence of reflections with other obstacles present in the corridor rather than walls (e.g. windows, cabinets or doors). 

The intercept parameter for $d$\:$=$\:1\:m is presented in Fig.~\ref{fig:fitting}(b). The differences between the empirical results and the $N$-rays model are in the order of 1\:dB, and both of them provides additional losses in the order of 3\:dB respect to the expected free-space values. This is a direct consequence of the optimal fitting with the power law model \eqref{eq:fitting}. Nevertheless, a decreasing trend is observed in both cases obtaining differences in path-gain between 250\:GHz and 330\:GHz of about 2\:dB. Finally, the minimum standard deviation is presented in Fig.\:\ref{fig:fitting}(c) at each frequency point for both empirical and $N$-rays model results. The mean values are 2.8\:dB and 2.5\:dB respectively, which imply a slightly higher variability in the empirical results. This is due to the fact that the corridors are not perfectly flat in their walls. Additional obstacles such as doors, intersections or cabinets might contribute to higher dispersion in path gain. 

The obtained values for the loss exponent (lower than the typical value of 2) demonstrates the suitability of high-rates point-to-point links at sub-THz frequencies within corridors. The free-space propagation losses are mitigated by the guiding effects of the corridor providing higher power at the receiver, extending the maximum range of the wireless system.

\subsection{Small Scale Effects}
The small scale effects are due to the MPCs present in the channel \cite{ref:Molisch2011}. These MPCs can be identified by analyzing the power delay profile (PDP), which is related to the CTF through the inverse discrete Fourier transform (IDFT). In particular, denoting $H[k,d]$\:$=$\:$H(f_k,d)$ as the CTF for the $k$-th frequency point and Tx-Rx distance $d$, the PDP is computed as
\begin{equation}
    P[n,d] = \left|\frac{1}{N}\sum_{k=0}^{N-1}W[k]H[k,d]\exp\left(j\frac{2\pi kn}{N}\right)\right|^2,
\end{equation}
where $n$\:$\in$\:$\{0,\dots,N-1\}$ is the iteration variable in the delay domain. This $n$ is related to both MPCs' delay $\tau_n$\:$=$\:$n/\mathrm{BW}$ and traveled distance $d_{\tau_n}$\:$=$\:$c\tau_n$. Thus, the delay distance resolution is $\Delta d_{\tau_n}$\:$=$\:3.75\:mm. Then, as mentioned in the previous section, the maximum observation distance in the delay domain is $d_{\max}$\:$=$\:$c/\Delta f$\:$=$\:30\:m. Lastly, a Hann window~$W[k]$ is applied through the frequency domain to reduce the sidelobes of sinc functions in the delay domain \cite{ref:Harris1978}.

\begin{figure}[t!]
    \centering
    \subfigure[CITIC-UGR.]{\includegraphics{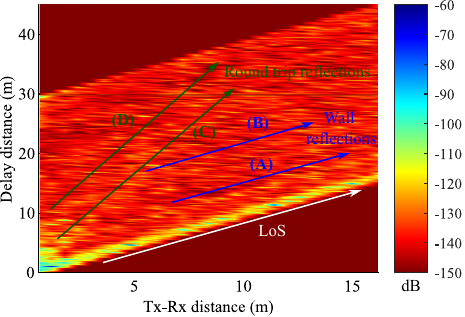}} \par\smallskip
    \subfigure[CETIC-UGR.]{\includegraphics{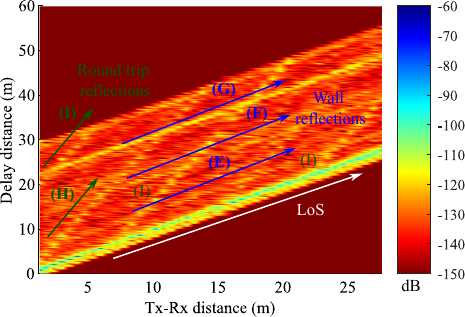}}
    \caption{Experimental PDP (in decibels) estimated from the measured CTFs in CITIC-UGR and CETIC-UGR. The identified MPCs are marked with an arrow and named with a letter of the alphabet.}
    \label{fig:pdp_surf}
\end{figure}

The experimental PDPs are depicted in Fig.\:\ref{fig:pdp_surf} for all \mbox{Tx-Rx} distances as heatmaps. This representation is expanded twice the maximum observation distance in delay domain $d_{\rm max}$\:$=$\:$c/\Delta f$\:$=$\:30\:m taking advantage of the LoS contribution (marked in white). Every component observed at delay distances shorter than the LoS are necessary aliased contributions that has traveled, at least, a distance longer than 30\:m. Therefore, as a consequence of the periodicity of the IDFT \cite{ref:Harris1978}, this region can be represented at delay distances in the range from 30\:m to 60\:m doubling the observation range.

\begin{figure}[t!]
    \centering
    \includegraphics{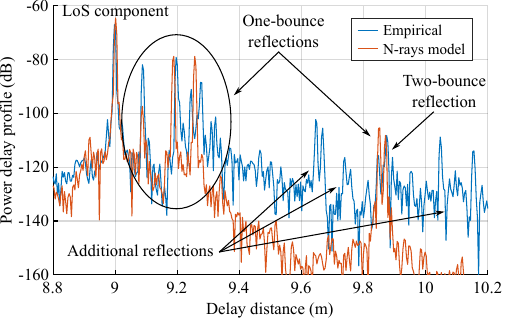}
    \caption{Comparison between empirical PDP and replication with $N$-rays model for a Tx-Rx distance $d$\:$=$\:9\:m in CITIC-UGR.}
    \label{fig:pdp_single}
\end{figure}

In both CITIC-UGR and CETIC-UGR, the LoS component corresponds with the line where the traveled distance in delay domain is equal to the Tx-Rx distance and provides the highest relative power. Additionally, two different types of MPCs are identified: (i)~reflections with the walls, marked in blue; and (ii) round trip reflections, marked in green. Reflections with the walls are easily identified as those lines parallels to the LoS contribution. In fact, the expression \eqref{eq:reflections_distance} for $d_n^\pm$ can be approximated as the \mbox{Tx-Rx} distance plus a certain constant. For the CITIC-UGR corridor [see Fig.\:\ref{fig:pdp_surf}(a)], up to two reflections, denoted by (A) and (B), are clearly identified distanced from the LoS component 4\:m and 8.4\:m respectively. For the CETIC-UGR corridor [see Fig.\:\ref{fig:pdp_surf}(b)], three different reflections are observed, denoted by (E), (F) and (G). These contributions are distanced to the LoS component 5\:m, 12\:m and 22\:m respectively. As a consequence, the reflections (F) and (G) appears in the extended region when their delay distance is larger than 30\:m. In addition to these contributions, there are up to 4 reflected rays distanced from the LoS component up to 0.4\:m. These MPCs corresponds with the one-bounce wall reflections, whose traveled distance is the closest to the LoS component [see eq. \eqref{eq:reflections_distance}]. An example for a Tx-Rx distance $d$\:$=$\:9\:m in CITIC-UGR is presented in Fig.\:\ref{fig:pdp_single} comparing both empirical and $N$-rays model PDPs. Additional reflections are observed in the empirical PDP along with the reflections with the walls.

Finally, the round trip reflections are identified as those that traveled towards to the Rx, then were reflected back to the Tx and, lastly, were reflected again to the Rx. Therefore, the distance traveled by these waves is in the order of 3$d$; that is, the slope of the line traced in the PDP diagrams is three times the one observed for the LoS contribution. In CITIC-UGR corridor [see Fig.\:\ref{fig:pdp_surf}(a)] are observed two of these MPCs denoted by (C) and (D). The former corresponds with a direct reflection with the positioners of Tx and Rx, and the latter has traveled bouncing with the walls of the corridor. These are analogous to the contributions (H) and (I) in CETIC-UGR corridor [see Fig.\:\ref{fig:pdp_surf}(b)]. Nevertheless, in this case the relative power is high enough to observe them in the aliasing region. Moreover, because of the larger slope, the aliased contributions extend to distances larger than the LoS contribution.

The previous analysis reveals that corridors are environments with a notable multipath effect. This contributes to the appearance of small-scale or fast fading effects which can notably degrade the transmitted signals across the wireless channel. In particular, these can be characterized by studying parameters such as the $K$-factor, the RMS delay spread $D_s$ and the coherence bandwidth $B_c$ \cite{ref:Molisch2011}.

The $K$-factor accounts for the increment in power of the main contribution (generally the LoS ray) respect to other components \cite{ref:Durgin2002}. Wireless systems like single-input single-output (SISO) requires of an strong contribution to operate properly, which is translated into high values of $K$. In that cases, the fast fading effects are not relevant and the channel is dominated by the main contribution. Conversely, in multiple-input multiple-output (MIMO) systems, this \mbox{$K$-factor} is desired to be low; that is, any contribution stands out respect to each other. In that case, the small scale effects are pronounced contributing to high decorrelation between the antennas which translates into higher capacities \cite{ref:Molisch2011}.

\begin{figure}[t!]
    \centering
    \subfigure[CITIC-UGR.]{\includegraphics{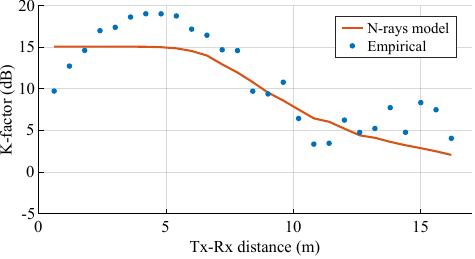}} \par\smallskip
    \subfigure[CETIC-UGR.]{\includegraphics{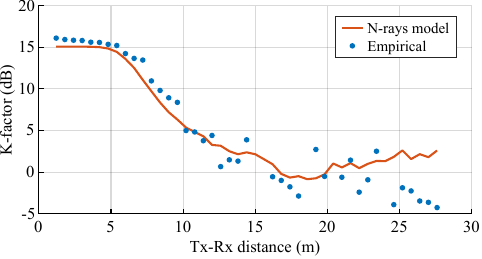}}
    \caption{Estimated $K$-factor for each Tx-Rx distance in both CITIC-UGR and CETIC-UGR corridors.}
    \label{fig:k-factor}
\end{figure}

One of the most popular approaches to estimate the \mbox{$K$-factor} from the CTF is the well-known method of moments (MoM) \cite{ref:Greenstein1999}. In particular, the estimated value of $K$ is given by
\begin{equation}
    \hat{K} = \frac{\sqrt{G_a^2 - G_v}}{G_a - \sqrt{G_a^2 - G_v}},
\end{equation}
where $G_a$ and $G_v$ are the second and forth order moments of the frequency response:
\begin{equation}
    G_a = \frac{1}{N}\sum_{k=1}^N|H[k]|^2,
\end{equation}
\begin{equation}
    G_v = \frac{1}{N-1}\left(\sum_{k=1}^N|H[k]|^4 - NG_a^2\right).
\end{equation}
The obtained results for both CITIC-UGR and CETIC-UGR are depicted in Fig.\:\ref{fig:k-factor}, comparing the empirical samples with the $N$-rays model behavior.

In CITIC-UGR corridor [see Fig.\:\ref{fig:k-factor}(a)], the $K$-factor is higher than 10\:dB for distances $d$\:$<$\:7.2\:m for both empirical samples and $N$-rays model. This is due to the fact that the LoS component is stronger than any other reflection. However, clear differences in trend are observed between the measurements and the $N$-rays model replication. Whereas the latter exhibits a constant value of 15\:dB, the experimental $K$-factor follows an increasing trend due to the presence of the strong round trip reflections previously identified in the PDP analysis. For distances $d$\:$\geq$\:7.2\:m, the $K$-factor decreases in both cases with a similar trend as the LoS component is attenuated. Similarly, in CETIC-UGR corridor [see Fig.\:\ref{fig:k-factor}(b)], the $K$-factor remains constant at 15\:dB for $d$\:$<$\:7.2\:m. For 7.2\:m\:$\leq$\:$d$\:$<$\:16.8\:m, the value of $K$ starts to decay down to 0\:dB. Finally, for $d$\:$\geq$\:16.8\:m, the empirical $K$-factor oscillates around this constant. This variability is due to the randomness of the wireless channel and also explains the difference with the $N$-ray model, which exhibits an slight increasing trend because of the attenuation of the LoS component compared to other reflections. In summary, the obtained $K$-factor ranges values from approximately 0\:dB to 15\:dB, which is line with other works at mmWave and sub-THz bands \cite{ref:Yue2019,ref:Takahashi2025}.

The RMS delay spread is a condensed parameter of the wireless channel that accounts for the spread of the different MPCs. The higher $D_s$ is, the larger is the spread of the components along the delay domain. This parameter can be estimated weighting with the PDP as follows: 
\begin{equation}
    D_s = \sqrt{\frac{\sum_{n=0}^{N-1}P[n](\tau_n -\bar{\tau})^2}{\sum_{n=0}^{N-1}P[n]}},
\end{equation}
where $\bar{\tau}$ represents the mean delay and is obtained as
\begin{equation}
    \bar{\tau}=\frac{\sum_{n=0}^{N-1}P[n]\tau_n}{{\sum_{n=0}^{N-1}P[n]}}.
\end{equation}
In addition, for experimental analysis, a certain threshold relative to the maximum power value is applied to the PDP before computing $D_s$ in order to mitigate noise effects. For this work, a typical value of 20\:dB has been set as a trade-off between lost components and avoided noise. A higher threshold (e.g. 30\:dB) provides noisy results with no clear trend. In contrast, a lower threshold (e.g. 10\:dB) underestimate delay spread obtaining low values for large Tx-Rx distances.

\begin{figure}[t!]
    \centering
    \subfigure[CITIC-UGR.]{\includegraphics{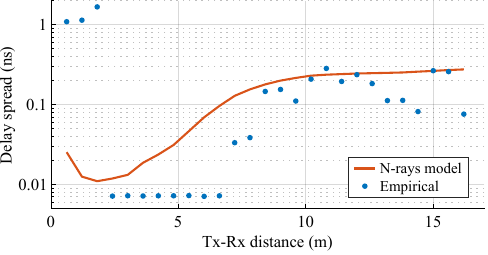}} \par\smallskip
    \subfigure[CETIC-UGR.]{\includegraphics{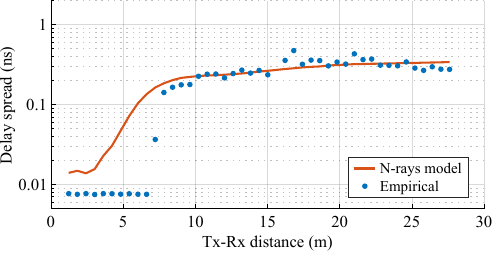}}
    \caption{Estimated delay spread as a function of Tx-Rx distance in both CITIC-UGR and CETIC-UGR corridors.}
    \label{fig:delay-spread}
\end{figure}

All the results are presented in Fig.\:\ref{fig:delay-spread} for both CITIC-UGR and CETIC-UGR, where similar trends are observed. The behavior is opposite to the $K$-factor: for distances $d$\:$<$\:7.2\:m, the delay spread is close to zero due to the high power of the LoS contribution compared to the rest of MPCs whereas for $d$\:$\geq$\:7.2\:m increases as MPCs power become comparable to the LoS. This trend, observed in the empirical samples, is properly captured by the $N$-rays model, which has been computed without thresholding the PDPs because the absence of noise. Therefore, a smooth transition between the low and high $D_s$ regions is observed. Nevertheless, particularly for the CITIC-UGR corridor, at distances $d$\:$<$\:1.8\:m the empirical delay spread is higher than 1\:ns, reaching a maximum value of 1.6\:ns because of the presence of the strong round-trip reflections previously mentioned. These results are substantially lower than the observed at mmWave bands in works such as \cite{ref:Shen2021,ref:Yue2019,ref:Geng2009,ref:Maccartney2015}, which are in the range between 10\:ns and 20\:ns. Conversely, these results seems coherent with others obtained at sub-THz bands for directive antennas \cite{ref:Wang2025,ref:Takahashi2025,ref:Lee2024}, in the order of 1\:ns and below. Therefore, the empirical results demonstrate that RMS delay spread is substantially lower in the THz bands than the expected at mmWave.

\begin{figure}[t!]
    \subfigure[CITIC-UGR.]{\includegraphics{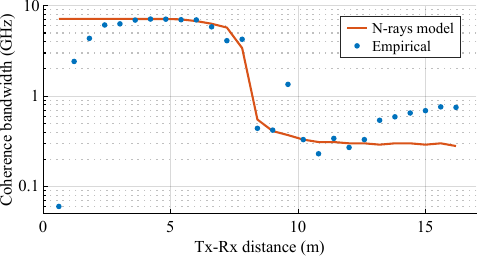}} \par\smallskip
    \subfigure[CETIC-UGR.]{\includegraphics{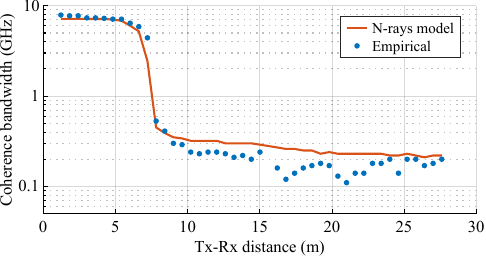}}
    \caption{Estimated coherence bandwidth for both CITIC-UGR and CETIC-UGR corridors at each Tx-Rx distance.}
    \label{fig:coherence-bandwidth}
\end{figure}

Lastly, the coherence bandwidth can be interpreted as the maximum bandwidth where the channel can be considered frequency-flat. This is strongly related to the $K$-factor and RMS delay spread since is completely dependent on the multipath effect. More specifically, it is a parameter that accounts for the frequency selectivity of a channel and is obtained from the frequency correlation function (FCF). This spread function can be estimated from the measured CTF as
\begin{equation}
    \begin{array}{rl}
        \displaystyle R_H[\Delta k] = \sum_{k=0}^{N-\Delta k-1}H[k + \Delta k]H^*[k], & \Delta k\geq0.
    \end{array}
\end{equation}
The values for $\Delta k$\:$<$\:0 are obtained as $R_H[\Delta k]$\:$=$\:$R_H^*[-\Delta k]$. Usually, it is normalized to its maximum:
\begin{equation}
    \hat{R}_H[\Delta k] = \frac{R_H[\Delta k]}{R_H[0]}
\end{equation}
Then, the coherence bandwidth is defined as the frequency difference required for a correlation coefficient lower than a certain threshold. Particularly, in this work a threshold value of 0.9 has been set obtaining the results depicted in Fig.\:\ref{fig:coherence-bandwidth} for both CITIC-UGR and CETIC-UGR.

In this case, the trend is analogous to the one observed for the $K$-factor and inverse to the delay spread. At distances $d$\:$<$\:7.2\:m, the coherence bandwidth is notably high (in the order of GHz) due the strong power of the LoS component respect to the rest of MPCs. However, in the CITIC-UGR corridor, the strong round-trip reflections decreases the value. In contrast, for distances $d$\:$\geq$\:7.2\:m, the coherence bandwidth decreases to the order of hundreds of MHz because the wall reflections become relevant. As for the other parameters, the $N$-rays model follows the empirical trend properly giving an accurate method to evaluate the propagation behavior within indoor corridors.

The values obtained for $K$, $D_s$ and $B_c$ are coherent with the presence of the reflected components within the corridor at distances higher than 7.2\:m. Nevertheless, it can be also inferred that their influence on the received signal is limited by the power of the LoS contribution. This fact explains the similarity between the $K$-factor values obtained with other works at mmWave bands but the notable differences in delay spread, indicating a dominance of the LoS component and higher stability within the frequency band. Therefore, as mentioned in the previous sub-section, these sub-THz link are suitable to deploy point-to-point links, minimizing the interference effects between MPCs.

Finally, all the numerical results of both large and small scale parameters are summarized in Tables \ref{tab:large_scale} and \ref{tab:small_scale} respectively. From the former, a mean path gain exponent of $-$1.8 along the 80\:GHz bandwidth is obtained, which is tied to a guiding effect of the corridor consequence of the reflections with walls, floor and ceil. Moreover, a mean standard deviation of 2.8\:dB which is aligned with other LoS campaigns within corridors at lower bands. On the other hand, from Table \ref{tab:small_scale} is shown that for short Tx-Rx distances the $K$-factor values are higher than 10\:dB, whereas it reduces to values in the order of 0\:dB as distance increases. These results are aligned with other works conducted in mmWave bands. On the other hand, the coherence bandwidth changes from values in the order of 6\:GHz for short distances to 100\:MHz when Tx-Rx separation increases. Conversely, the delay-spread increases with Tx-Rx distance from values in the order of 0.01\:ns to 0.2\:ns, which are notably lower than the observed at mmWave bands. 

\begin{table*}[t!]
    \centering
    \begin{minipage}{0.45\linewidth}
        \centering
        \caption{Numerical results obtained from empirical path gain modeling with power law}
        \label{tab:large_scale}
        \renewcommand{\arraystretch}{1.25}
        \begin{tabular}{|c||c|c|c|}
            \hline
            \textbf{Frequency [GHz]} & $n(f)$ & $A(f)$ [dB] & $\sigma(f)$ [dB] \\ \hline \hline
            250 & $-$1.74 & $-$83.1 & 3.6 \\ \hline
            255 & $-$1.76 & $-$83.1 & 3.1 \\ \hline
            260 & $-$1.80 & $-$82.6 & 2.8 \\ \hline
            265 & $-$1.62 & $-$84.2 & 3.5 \\ \hline
            270 & $-$1.81 & $-$84.3 & 2.9 \\ \hline
            275 & $-$1.86 & $-$83.4 & 2.9 \\ \hline
            280 & $-$1.73 & $-$84.4 & 3.0 \\ \hline
            285 & $-$1.68 & $-$85.0 & 2.7 \\ \hline
            290 & $-$1.80 & $-$84.8 & 3.4 \\ \hline
            295 & $-$1.76 & $-$85.0 & 2.3 \\ \hline
            300 & $-$2.00 & $-$83.2 & 2.5 \\ \hline
            305 & $-$1.90 & $-$84.4 & 3.0 \\ \hline
            310 & $-$1.84 & $-$85.6 & 2.4 \\ \hline
            315 & $-$2.01 & $-$84.9 & 2.3 \\ \hline
            320 & $-$1.81 & $-$86.5 & 2.5 \\ \hline
            325 & $-$1.89 & $-$86.3 & 2.5 \\ \hline
            330 & $-$1.99 & $-$85.8 & 2.7 \\ \hline
        \end{tabular}
    \end{minipage}
    \hspace{0.05\linewidth}
    \begin{minipage}{0.45\linewidth}
        \centering
        \caption{Small-scale parameters values given in the form (CITIC-UGR / CETIC-UGR) for each Tx-Rx distance}
        \label{tab:small_scale}
        \renewcommand{\arraystretch}{1.25}
        \begin{tabular}{|c||c|c|c|}
            \hline
            \textbf{Distance [m]} & $K$ [dB] & $D_s$ [ns] & $B_c$ [MHz] \\ \hline \hline
            0.6 & 9.7 / - & 1.09 / - & 60 / - \\ \hline
            2.4 & 17.0 / 15.9 & 0.01 / 0.01 & 6130 / 7750 \\ \hline
            3.0 & 17.4 / 15.8 & 0.01 / 0.01 & 6310 / 7350 \\ \hline
            4.2 & 19.0 / 15.6 & 0.01 / 0.01 & 7130 / 7260 \\ \hline
            6.0 & 17.2 / 14.2 & 0.01 / 0.01 & 6980 / 6420 \\ \hline
            7.2 & 14.7 / 13.5 & 0.03 / 0.04 & 4110 / 4420 \\ \hline
            7.8 & 14.6 / 11.0 & 0.04 / 0.14 & 4260 / 530 \\ \hline
            9.6 & 10.8 / 8.4 & 0.11 / 0.18 & 1350 / 290 \\ \hline
            11.4 & 3.5 / 3.8 & 0.20 / 0.24 & 340 / 240 \\ \hline
            13.2 & 5.2 / 1.5 & 0.11 / 0.27 & 540 / 210 \\ \hline
            15.0 & 8.4 / $-$6.0 & 0.27 / 0.24 & 690 / 240 \\ \hline
            17.4 & - / $-$1.8 & - / 0.32 & - / 140 \\ \hline
            19.2 & - / $-$2.7 & - / 0.31 & - / 180 \\ \hline
            21.0 & - / $-$0.6 & - / 0.43 & - / 110 \\ \hline
            22.8 & - / $-$0.9 & - / 0.31 & - / 180 \\ \hline
            24.6 & - / $-$3.9 & - / 0.34 & - / 140 \\ \hline
            26.4 & - / $-$3.5 & - / 0.30 & - / 170 \\ \hline
        \end{tabular}
    \end{minipage}

\end{table*}

\section{Conclusion}
Evaluation of THz channels in different scenarios is essential to properly design the next generation wireless systems. This work presents an extensive measurement campaign conducted in two different corridors at the H-band range from 250\:GHz to 330\:GHz. To the best of the authors' knowledge, for the first time in literature an UWB range of 80\:GHz is fully studied and modeled within corridors. Particularly, from the measured CTFs large and small scale effects are assessed. Path gain is modeled with a power law for all the frequencies in the band, obtaining variations in the range from $-$2.1 to $-$1.6. This behavior is also observed at mmWave bands and is attributed to a guiding effect of the corridor. Parameters such as $K$-factor, RMS delay spread and coherence bandwidth are evaluated and compared to mmWave bands results. Whereas $K$-factor distributes between 0\:dB and 15\:dB in both frequency ranges, delay spread is an order of magnitude lower than in mmWave bands ranging up to 1.6\:ns. Consequently, the obtained coherence bandwidth is particularly high with values in the order of hundreds of megahertz. Finally, in addition to the empirical assessment, an analytical approximation based on the $N$-rays model is proposed. This theoretical model accurately approaches to the wireless channel behavior in both large and small scale effects emerging as an useful tool to evaluate these scenarios. 

Overall, this work provides solid experimental results along with an accurate model which can be used jointly to enable the deployment of future 6G wireless systems with large bandwidth requirements such as massive data centers. As future work, additional nLoS measurements can be conducted to evaluate the propagation around corners to rooms or other corridors. Moreover, direction of arrival studies can be also performed in order to fully characterize the MPCs.

\section*{Appendix A \\ Frequency-Domain Averaging Criteria}
The Lee method for removing fast fading from a measurement consists on applying a moving average in distance with a window size between 20$\lambda$ and 40$\lambda$ \cite{ref:Lee1985}. For the sub-THz range, wavelengths are equal or lower than 1\:mm which imply displacements between measurements in the order of 10\:mm. That is not always possible because of technical or operational limitations, so an alternative is needed. In particular, for UWB measurements a moving average in frequency can be applied with an equivalent criteria \cite{ref:Molisch2011}.

In general, the electrical distance traveled by the LoS contribution varies linearly with frequency. Specifically, given a frequency $f$ and a Tx-Rx distance $d$, the electrical distance $d_\lambda$ is defined as
\begin{equation}
    d_\lambda = \frac{d}{\lambda} = \frac{d}{c}f.
\end{equation}
Therefore, for a window defined between frequencies $f_1$ and $f_2$ (with $f_2$\:$>$\:$f_1$), the difference on the traveled electrical distance is obtained as
\begin{equation}
    \Delta d_\lambda = d_{\lambda_2} - d_{\lambda_1} = \frac{d}{c}(f_2 - f_1).
\end{equation}
On the other hand, considering the Lee method criteria, this difference has to be equal to a natural number $M$ between 20 and 40, that is, $\Delta d_\lambda$\:$=$\:$M$. Then, the appropriate window size in frequency $L_f(d)$\:$=$\:$f_2-f_1$ is given by
\begin{equation}
    L_f(d) = \frac{Mc}{d}.
\end{equation}
This can be translated to number of frequency samples $L(d)$ by considering the frequency spacing $\Delta f$:
\begin{equation}
    L(d) = M\left\lceil\frac{c}{(\Delta f)d}\right\rceil + 1.
\end{equation}
Finally, the expression can be related to the maximum distance of the delay domain $d_{\rm max}$\:$=$\:$c/\Delta f$, which leads to
\begin{equation}
    L(d) = M\left\lceil\frac{d_{\rm max}}{d}\right\rceil + 1.
\end{equation}
This expression implies that the window size is lower as $d$ increases respect to the maximum excess distance in the delay domain. Therefore, the window size adapts to the coherence bandwidth of the channel since for larger distances additional MPCs are expected.

\end{document}

%% file: figures/channel_sounder.tex
\usetikzlibrary{shapes.symbols}
\node[cloud, draw =black, aspect=1.5, cloud puffs = 16, text width=2cm, align=center,scale=0.7] (mycloud) at (0,-3.5) {};
\node[align=center,text width=2cm] at (0,-3.5) {\footnotesize Propagation Channel};

\draw[fill=white] (-2.5,-1.5) rectangle (2.5,1.5) node[pos=0.5,text width=4cm,align=center] {\footnotesize \RS ZVA67 \\[-0.6ex] $[$13.89, 18.33$]$\:GHz};
\draw (-2.25,-0.75) rectangle (2.25,1.25);
\fill[gray,opacity=0.25] (-2.25,-0.75) rectangle (2.25,1.25);

\draw[stealth-] (-2.5,1) -- (-3,1) -- (-3,2.5) -- (-2.5,2.5);
\node[rotate=90] at (-3.25,1.75) {\footnotesize CLK REF (10\:MHz)};

\draw (-2.5,1.75) rectangle (2.5,3.5) node[pos=0.65,text width=4cm,align=center] {\footnotesize \RS SMF100A \\[-0.6ex] $[$10.41, 13.74$]$\:GHz};
\draw (-2.25,2) rectangle (-1,3);
\fill[gray,opacity=0.25] (-2.25,2) rectangle (-1,3);
\draw (2.125,2.125) circle (0.2);
\draw[*-stealth] (2.125-0.11,2.125) -- (3.5,2.125);
\draw (3.5,2.125-0.25) -- (3.5,2.125+0.25) -- (4,2.125+0.5) -- (4,2.125-0.5) -- (3.5,2.125-0.25);
\draw (3.75,2.125-1) node[text width=4cm,align=center] {\footnotesize Power \\[-0.6ex] divider}; 

\draw (-2.5,-0.75) -- (2.5,-0.75);

\draw (-1.25,-0.75) -- (-1.25,-1.5);
\draw (0,-0.75) -- (0,-1.5);
\draw (1.25,-0.75) -- (1.25,-1.5);

\draw (-2.125,-1.125) circle (0.2) node (P3) {};
\draw (-2.125+1.25,-1.125) circle (0.2) node (P1) {};
\draw (-2.125+2*1.25,-1.125) circle (0.2) node (P4) {};
\draw (-2.125+3*1.25,-1.125) circle (0.2) node (P2) {};

\draw (-1.75,-1.125) circle (0.1) node (R3) {};
\draw (-1.525,-1.125) circle (0.1) node (M3) {};
\draw[fill=black] (-1.75+1.25,-1.125) circle (0.1) node (R1) {};
\draw (-1.525+1.25,-1.125) circle (0.1) node (M1) {};
\draw (-1.75+1.25*2,-1.125) circle (0.1) node (R4) {};
\draw (-1.525+1.25*2,-1.125) circle (0.1) node (M4) {};
\draw (-1.75+1.25*3,-1.125) circle (0.1) node (R2) {};
\draw[fill=black] (-1.525+1.25*3,-1.125) circle (0.1) node (M2) {};

\draw (-4.5,-4) rectangle (-2.5,-3) node[pos=0.5,text width=2.25cm,align=center] {\footnotesize \RS \\[-0.6ex]  ZC330};
\node[] at (-2.125,-3.825) {\footnotesize Tx};
\draw[stealth-,dashed] (-4.5,-3.1-3*0.8/3) -- (-5,-3.1-3*0.8/3) node (TX_RF) {};
% \draw[-o] (-4.5,-3.1-2*0.8/3) -- (-5,-3.1-2*0.8/3) node (TX_MEAS) {};
\draw[-stealth,dotted] (-4.5,-3.1-0.8/3) -- (-5,-3.1-0.8/3) node (TX_REF) {};
\draw[stealth-] (-4.5,-3.1) -- (-5,-3.1) node (TX_LO) {};

\draw (4.5,-4) rectangle (2.5,-3) node[pos=0.5,text width=2.25cm,align=center] {\footnotesize \RS \\[-0.6ex] ZC330};
\node[] at (2.125,-3.825) {\footnotesize Rx};
\draw[stealth-] (4.5,-3.1-3*0.8/3) -- (5,-3.1-3*0.8/3) node (RX_LO) {};
% \draw[-o] (4.5,-3.1-2*0.8/3) -- (5,-3.1-2*0.8/3) node (RX_REF) {};
\draw[dotted,-stealth] (4.5,-3.1-0.8/3) -- (5,-3.1-0.8/3) node (RX_MEAS) {};
% \draw[stealth-o] (4.5,-3.1) -- (5,-3.1) node (RX_RF) {};

\draw (-5,-2) -- (-5.2,-2) -- (-5,-2.5) -- (-4.8,-2) -- (-5,-2);
% \draw (-5.45,-2.25) node[rotate=90] {\footnotesize 25\:dB};

\draw (4,2.125+0.25) -- (5.75,2.125+0.25) -- (5.75,3.75) -| (-5,-2);
\draw (-5,-2.5) -- (-5,-3.1);

\draw[dashed,*-] (-2.125+1.25,-1.125+0.11) |- (-4.25,-2) |- (-5.5,-2.75) |- (-4.5,-3.1-3*0.8/3) node[pos=0.5,below] {\footnotesize RF};
\draw[fill=white] (-3,-2) -- (-3,-1.8) -- (-3.5,-2) -- (-3,-2.2) -- (-3,-2);
% \draw (-3.25,-2.45) node[] {\footnotesize 25\:dB};

\draw[dotted] (-1.75+1.25,-1.125+0.0825) |- (-6,-1.675) |- (-5,-3.1-0.8/3);
% \draw (-6,-1.66)  node[above] {\footnotesize REF};
\node[rotate=90] at (-6.25,-2.52) {\footnotesize REF (279\:MHz)};

\draw (4,2.125-0.25) -| (5.75,-2) |- (5,-3.1-3*0.8/3);
\draw (3,2.140) node[above] {\footnotesize LO};

%\draw[fill=white] (4,-2) -- (4,-2.2) -- (4.5,-2) -- (4,-1.8) -- (4,-2);
\draw[fill=white] (5+0.75,-2) -- (5.2+0.75,-2) -- (5+0.75,-2.5) -- (4.8+0.75,-2) -- (5+0.75,-2);
% \draw (5.25,-2.25) node[rotate=90] {\footnotesize 25\:dB};

\draw[dotted] (-1.525+1.25*3-0.0825,-1.125) -- (5.25,-1.125) |- (5,-3.1-0.8/3);
\draw (3.875,-1.25) node[above,text width=2.25cm,align=center] {\footnotesize MEAS \\[-0.6ex] (279\:MHz)};

\draw (-2.5,-3.5) -| (-2,-2.5);
\draw (-2,-2.75) -- (-2.1,-2.55) (-2,-2.75) -- (-1.9,-2.55);

\draw (2.5,-3.5) -| (2,-2.5);
\draw (2,-2.75) -- (2.1,-2.55) (2,-2.75) -- (1.9,-2.55);

\draw (0,-4.75) node[below] {\footnotesize $[$250, 330$]$\:GHz};
\draw[-Latex] (-3.5,-4.25) arc (180:360:3.5 and 1.825) node[pos=0.5,below] {\footnotesize $H(f)$\:$=$\:$S_{21}(f)$};

% \draw[o-o] (-6,-5.5) -- (-4.5,-5.5);
% \draw[fill=white] (-5.5,-5.3) -- (-5.5,-5.7) -- (-5,-5.5) -- (-5.5,-5.3); 
% \node[text width=1.25cm,align=center] at (-5.25,-6.25) {\footnotesize PA 2522 \\[-0.6ex] 25\:dB};
% \draw[dashed] (-6.125,-5.175) rectangle (-4.375,-6.75);

\draw[o-o] (-6,-5) -- (-4.5,-5);
\draw[fill=white] (-5.5,-4.8) -- (-5.5,-5.2) -- (-5,-5) -- (-5.5,-4.8); 
\node[text width=1.25cm,align=center] at (-5.25,-5.75) {\footnotesize PA 2522 \\[-0.6ex] 25\:dB};
\draw[dashed] (-6.125,-4.675) rectangle (-4.375,-6.25);

% \draw[o-] (4.25,-5.5) -- (4.75,-5.5);
% \draw[-o] (5.25,-5.25) -- (5.75,-5.25);
% \draw[-o] (5.25,-5.75) -- (5.75,-5.75);
% \draw[fill=white] (4.75,-5.75) -- (4.75,-5.25) -- (5.25,-5) -- (5.25,-6) -- (4.75,-5.75);
% \draw[fill=white] (3.5,2.125-0.25) -- (3.5,2.125+0.25) -- (4,2.125+0.5) -- (4,2.125-0.5) -- (3.5,2.125-0.25);
% \node[text width=1.25cm,align=center] at (5,-6.5) {\footnotesize \red{Model} \\[-0.6ex] $-$3\:dB};
% \draw[dashed] (5.875,-4.85) rectangle (4.125,-7);

%% file: figures/nray_model_odd.tex
\definecolor{MBblue}{rgb}{0.0660,0.4430,0.7450}
\definecolor{MBred}{rgb}{0.8660,0.3290,0.0}

% Corridor walls/ceiling-floor.
\draw[line width = 1.25pt](0,1.75) -- (8,1.75) (0,-1.75) -- (8,-1.75);
\node[] at (7.5,1.95) {\footnotesize Wall};

% Corridor central axis.
\draw[dotted] (0.125,0) -- (7.875,0);
\node[] at (6,-0.2) {\footnotesize Axis};

% LoS ray.
\draw[gray] (0.75,0.5) -- (7.25,0.5) node[currarrow,pos=0.125,xscale=0.75,sloped,scale=0.75,gray] {}
node[currarrow,pos=0.875,xscale=0.75,sloped,scale=0.75,gray] {}
node[currarrow,pos=0.425,xscale=0.75,sloped,scale=0.75,gray] {}
node[currarrow,pos=0.575,xscale=0.75,sloped,scale=0.75,gray] {};

% 3 bounces reflections.
\draw[MBblue] (0.75,0.5) -- (1.61,1.75) node[currarrow,pos=0.5,xscale=0.75,sloped,scale=0.75,MBblue] {} -- (4,-1.75) node[currarrow,pos=0.25,xscale=0.75,sloped,scale=0.75,MBblue] {}
node[currarrow,pos=0.75,xscale=0.75,sloped,scale=0.75,MBblue] {} -- (6.39,1.75) node[currarrow,pos=0.25,xscale=0.75,sloped,scale=0.75,MBblue] {}
node[currarrow,pos=0.75,xscale=0.75,sloped,scale=0.75,MBblue] {} -- (7.25,0.5) node[currarrow,pos=0.5,xscale=0.75,sloped,scale=0.75,MBblue] {};

\draw[MBred] (0.75,0.5) -- (2.02,-1.75) node[currarrow,pos=0.5,xscale=0.75,sloped,scale=0.75,MBred] {} -- (4,1.75) node[currarrow,pos=0.25,xscale=0.75,sloped,scale=0.75,MBred] {} node[currarrow,pos=0.75,xscale=0.75,sloped,scale=0.75,MBred] {} -- (5.98,-1.75) node[currarrow,pos=0.25,xscale=0.75,sloped,scale=0.75,MBred] {} node[currarrow,pos=0.75,xscale=0.75,sloped,scale=0.75,MBred] {} -- (7.25,0.5) node[currarrow,pos=0.5,xscale=0.75,sloped,scale=0.75,MBred] {};

% Tx and Rx positions.
\draw[fill=black] (0.75,0.5) circle (0.05) node[left,yshift=6.25] {\footnotesize Tx};
\draw[fill=black] (7.25,0.5) circle (0.05) node[right,yshift=6.25] {\footnotesize Rx};

% Distances.
\draw[dashed] 
    (0.75,0.5) -- (0.75,-2.875) 
    (7.25,0.5) -- (7.25,-2.875)
    (7.25,0.5) -- (8.0625,0.5)
    
    (1.61,1.75) -- (1.61,2.125)
    (2.02,-1.75) -- (2.02,-2.125)
    (4,-2.125) -- (4,2.125)
    (5.98,-1.75) -- (5.98,-2.125)
    (6.39,1.75) -- (6.39,2.125);

\draw (4,-1) arc (90:55.6197:0.75) node[pos=0.75,above] {\footnotesize $\theta_3^+$};
\draw (4,1) arc (270:240.5241:0.75) node[pos=0.625,below] {\footnotesize $\theta_3^-$};

\draw[stealth-stealth] 
    (0.75,-2.75) -- (7.25,-2.75) node[below,pos=0.5] {\footnotesize $d$};
\draw[stealth-stealth] 
    (0.0625,-1.75) -- (0.0625,1.75) node[left,pos=0.5] {\footnotesize $w$};
\draw[stealth-stealth] 
    (7.9375,0) -- (7.9375,0.5) node[right,pos=0.5] {\footnotesize $\Delta w$};

\draw[stealth-stealth] 
    (1.61,2) -- (4,2) node[above,pos=0.5] {\footnotesize $\delta_3^+$}; 
\draw[stealth-stealth] 
    (4,2) -- (6.39,2) node[above,pos=0.5] {\footnotesize $\delta_3^+$};

\draw[stealth-stealth] 
    (2.02,-2) -- (4,-2) node[below,pos=0.5] {\footnotesize $\delta_3^-$}; 
\draw[stealth-stealth] 
    (4,-2) -- (5.98,-2) node[below,pos=0.5] {\footnotesize $\delta_3^-$};

%% file: figures/nray_model_even.tex
\definecolor{MBblue}{rgb}{0.0660,0.4430,0.7450}
\definecolor{MBred}{rgb}{0.8660,0.3290,0.0}

% Corridor walls/ceiling-floor.
\draw[line width = 1.25pt] (0,1.75) -- (8,1.75) (0,-1.75) -- (8,-1.75);
\node[] at (7.5,1.95) {\footnotesize Wall};

% Corridor central axis.
\draw[dotted] (0.125,0) -- (7.875,0);
\node[] at (2.5,-0.2) {\footnotesize Axis};

% LoS ray.
\draw[gray] (0.75,0.5) -- (7.25,0.5) node[currarrow,pos=0.125,xscale=0.75,sloped,scale=0.75,gray] {}
node[currarrow,pos=0.875,xscale=0.75,sloped,scale=0.75,gray] {}
node[currarrow,pos=0.425,xscale=0.75,sloped,scale=0.75,gray] {}
node[currarrow,pos=0.575,xscale=0.75,sloped,scale=0.75,gray] {};

% 2 bounces reflections.
\draw[MBblue] (0.75,0.5) -- (1.91,1.75) 
node[currarrow,pos=0.5,xscale=0.75,sloped,scale=0.75,MBblue] {} -- (5.16,-1.75) 
node[currarrow,pos=0.25,xscale=0.75,sloped,scale=0.75,MBblue] {} 
node[currarrow,pos=0.75,xscale=0.75,sloped,scale=0.75,MBblue] {} -- (7.25,0.5)
node[currarrow,pos=0.5,xscale=0.75,sloped,scale=0.75,MBblue] {};

\draw[MBred] (0.75,0.5) -- (2.84,-1.75)
node[currarrow,pos=0.5,xscale=0.75,sloped,scale=0.75,MBred] {} -- (6.09,1.75)
node[currarrow,pos=0.25,xscale=0.75,sloped,scale=0.75,MBred] {} 
node[currarrow,pos=0.75,xscale=0.75,sloped,scale=0.75,MBred] {} -- (7.25,0.5)
node[currarrow,pos=0.5,xscale=0.75,sloped,scale=0.75,MBred] {};

% Tx and Rx positions.
\draw[fill=black] (0.75,0.5) circle (0.05) node[left,yshift=6.25] {\footnotesize Tx};
\draw[fill=black] (7.25,0.5) circle (0.05) node[right,yshift=6.25] {\footnotesize Rx};

% Distances.
\draw[dashed] 
    (0.75,0.5) -- (0.75,-2.875) 
    (7.25,0.5) -- (7.25,-2.875)
    (7.25,0.5) -- (8.0625,0.5)
    
    (1.91,1.75) -- (1.91,2.125)
    (2.84,-1.75) -- (2.84,-2.125)
    (5.16,-1.75) -- (5.16,2.125)
    (6.09,1.75) -- (6.09,-2.125);

\draw (5.16,-1) arc (90:47.1211:0.75) node[pos=0.75,above] {\footnotesize $\theta_2^+$};
\draw (6.09,1) arc (270:227.1211:0.75) node[pos=0.625,below] {\footnotesize $\theta_2^-$};

\draw[stealth-stealth] 
    (0.75,-2.75) -- (7.25,-2.75) node[below,pos=0.5] {\footnotesize $d$};
\draw[stealth-stealth] 
    (0.0625,-1.75) -- (0.0625,1.75) node[left,pos=0.5] {\footnotesize $w$};
\draw[stealth-stealth] 
    (7.9375,0) -- (7.9375,0.5) node[right,pos=0.5] {\footnotesize $\Delta w$};

\draw[stealth-stealth] 
    (1.91,2) -- (5.16,2) node[above,pos=0.5] {\footnotesize $\delta_2^+$}; 

\draw[stealth-stealth] 
    (2.84,-2) -- (6.09,-2) node[below,pos=0.5] {\footnotesize $\delta_2^-$}; 